\documentclass[]{revtex4}
\usepackage{graphicx}
\usepackage{epstopdf}
\usepackage{color}

\newcommand{\DEL}[1]{}
\newcommand{\ADD}[1]{#1}
\newcommand{\fref}[1]{ figure~\ref{#1}}
\newcommand{\eref}[1]{ equation~\ref{#1}}
\newcommand{\Fref}[1]{ Figure~\ref{#1}}

\begin{document}
\title{Influence of Reynolds number and forcing type in a turbulent  von K\'arm\'an flow}

\author{B. Saint-Michel$^1$ \footnote{Present affiliation : Aix Marseille Universit\'e, CNRS, Centrale Marseille, IRPHE UMR 7342, 13384, Marseille, France}, B. Dubrulle$^1$, L. Mari\'e$^2$, 
F. Ravelet$^3$, F.~Daviaud$^1$}
\affiliation{$^1$ Laboratoire SPHYNX, Service de Physique de l'\'Etat Condens\'e, DSM, CEA Saclay, CNRS URA 2464, 91191 Gif-sur-Yvette, France}
\affiliation{$^2$ Laboratoire de Physique des Oc\'eans, UMR 6523 CNRS/IFREMER/IRD/UBO, Brest, France}
\affiliation{$^3$ Laboratoire Dynfluid, ENSAM ParisTech, CNRS EA92, 151, boulevard de l'H\^{o}pital 75013 Paris, France}
\email{brice.saint-michel@cea.fr}

\begin{abstract}
We present a detailed study of of a global bifurcation occuring in a turbulent von K\'arm\'an swirling flow. In this system, the statistically steady states progressively display hysteretic behaviour when the Reynolds number is increased above the transition to turbulence. We examine in detail this hysteresis using asymmetric forcing conditions --- rotating the impellers at different speeds. For very high Reynolds numbers, we study the sensitivity of the hysteresis cycle --- using complementary Particle Image Velocimetry (PIV) and global mechanical measurements --- to the forcing nature, imposing either the torque or the speed of the impellers. New mean states, displaying multiple quasi-steady states and negative differential responses, are experimentally observed in torque control. A simple analogy with electrical circuits is performed to understand the link between multi-stability and negative responses. The system is compared to other, similar ``bulk" systems, to understand some relevant ingredients of negative differential responses, and studied in the framework of thermodynamics of long-range interacting systems. \ADD{The experimental results are eventually compared to the related problem of Rayleigh-B\'enard turbulence.}
\end{abstract}

\maketitle

\section{Introduction}

Most of the flows present at human or natural scales are turbulent flows, for which the local (Eulerian) velocity fluctuations follow the Kolmogorov spectrum~\cite{Frisch1995}. The transition from laminar to turbulent flows is parametrised by the Reynolds number, a non-dimensional number comparing the inertial and the viscous terms in the Navier-Stokes equations governing the motion of fluid particles. Intense effort has focused on measuring (and predicting) the onset of turbulence in various geometries such as plane Couette~\cite{Daviaud1992,Orszag1980}, Taylor Couette~\cite{Lathrop1992}, channel~\cite{Orszag1980}, pipe~\cite{Reynolds1883,Pfenniger1961}, and cylinder wake~\cite{Bloor1964} flows.
 \ADD{In such flows, the global symmetries of the laminar flow (time and spatial invariances) are progressively broken for increasing Reynolds numbers (${\rm Re}$). For large enough ${\rm Re}$, all symmetries of the experimental set-up are broken in the instantaneous flow. Concurrently, a range of scales, the so-called ``inertial range" of turbulence, in which the flow fluctuations exhibit universal statistical properties, develops between the forcing scale and the Kolmogorov scale, the latter scale decreasing as the Reynolds number tends to infinity.}
 
\ADD{Modern turbulence modelling relies on stochastic processes to account for the very high level of ``noise" present in turbulent flows, and are designed to reproduce the statistical properties observed in experiments: in that respect, this often leads to an ergodic interpretation of the local flow~\cite{Frisch1995, Monin1975} where the temporal evolution of the physical quantities is statistically steady, and the noise level is sufficient to explore all the possible states of the turbulent flow. In this situation, experimental symmetries are recovered in a statistical sense at small scales, and away from the boundaries~\cite{Frisch1995}. Whether this restoration is also observed for the large scale flow is currently unknown but is a commonly accepted idea, at least for cylinder wake and pipe flows, despite being challenged for example by some turbulent Rayleigh-B\'enard experiments exhibiting, in some particular cases, a symmetry-breaking steady large-scale wind~\cite{Araujo2005}.}

 \DEL{In some cases, the transition to turbulence is subcritical, and the critical Reynolds number for turbulence onset $R_c$ is non universal and may depend on the experimental noise. However, it is generally assumed that for ${\rm Re}$ large enough, the flow is considered to be turbulent, and is characterized by  "universal" properties (statistical symmetry, independence of the geometry, forcing or dissipation) in a suitable range of scale called  ``inertial".} 
 
Additional light about this assumption has recently been shed by a series of experiments in a von K\'arm\'an flow geometry, with flow forced by counter-rotating disks fitted with blades. In that case, the progressive transition to turbulence occurs for Reynolds numbers $1000 \leq {\rm Re} \leq 3300$ (depending upon blade curvature)~\cite{Ravelet2008}. Past this value, the flow first reaches  statistically steady states restoring the symmetries of the experimental setup and with characteristics reminiscent of other turbulent flows (Kolmogorov spectrum for local velocity measurements).  However, for Reynolds numbers above ${\rm Re = 10\,000}$, the turbulent flow undergoes another, \emph{global} bifurcation, \ADD{in which the \emph{statistically steady} turbulent states undergo a bifurcation and become hysteretic, breaking at large scales the ergodic hypothesis~\cite{Ravelet2004,Ravelet2008}}. This confirms that the \DEL{concept of "fully turbulent" flow is not meaningful at large scales} \ADD{symmetry restoration principles are not always valid for the large-scale flow, or should at least be considered with great caution}. On the other side, this calls for supplementary studies of the large scale behaviours, to study properties of this special regime and investigate possible emergence of  universal features.

In this article, we focus on the properties of the hysteresis cycle of the global bifurcation by studying the response of the flow to a symmetry-breaking field for various Reynolds numbers lying between ${\rm Re} = 800$ ---~corresponding to a non-hysteretic, non turbulent flow~--- to ${\rm Re} = 250\,000$ ---~corresponding to a hysteretic flow which displays all the characteristics of turbulent flows~---. In addition, we  explore the influence of the \emph{nature} of the forcing conditions on the properties of the hysteresis cycle, and  on the set of states accessible to the experiment. We compare our experimental results to other out-of-equilibrium systems and we show that the peculiar large scale behaviour we observe mirrors characteristic properties of long-range interacting systems at thermodynamic equilibrium.

\section{Notations and experimental parameters}

\subsection{Mechanical setup}

\begin{figure}
	\centering
	\includegraphics[width = 0.60 \textwidth, trim = {0cm 0cm 9cm 0cm}, clip = true]{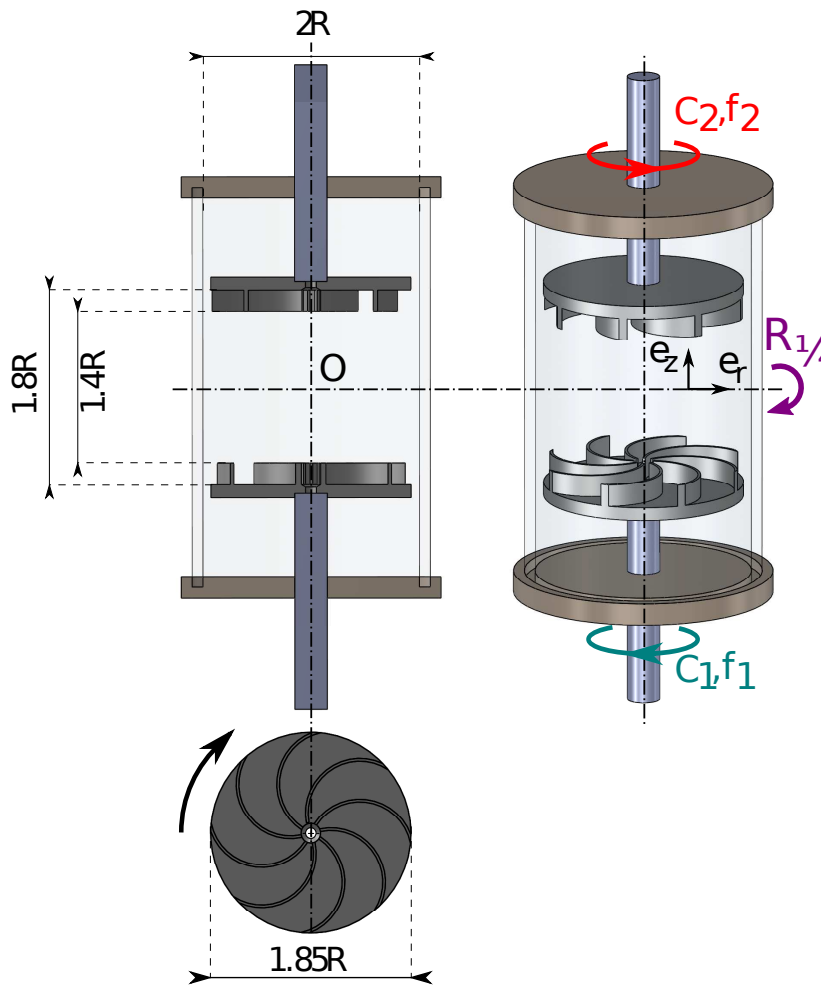}
	\caption{Experimental setup, with blade profile. The rotation sense defined by the arrow is called $(-)$.}
	\label{fig:setup}
\end{figure}

The experimental set-up used in the article is essentially the same as in~\cite{Cortet2011}: a Plexiglas cylinder of inner radius $R = 100$~mm is filled with fluid, which is stirred by two coaxial discs of radius $R = 0.925$~mm made of Polycarbonate (see \fref{fig:setup}). The discs, fitted with blades of curvature radius $0.4625 R$ and height $0.200 R$, are separated by a distance $1.4 R$ (from blade tip to blade tip) and are driven by two independent synchronous motors of nominal power $1.8$~kW. Motors can function either imposing the speed~\cite{Cortet2011} or the torque to the discs, in both directions of rotation, called $(+)$ and $(-)$. In this article, we will only consider (unless explicitly stated) experiments for which the two discs are counter-rotating and pushing the fluid with the concave face of the blades (clock-wise direction in \fref{fig:setup}) which defines direction $(-)$. 

Two sets of experiments have been studied in this article. The first set contains exclusively speed-control experiments in water-glycerol mixtures of various concentrations allowing a large span of fluid viscosities $\nu$ (and therefore, of Reynolds numbers) using impellers with 16 blades. Such experiments have used a lip seal to ensure fluid confinement in the experiment. All of the experiments of the second set have been performed in water, using \emph{both} controls (speed and torque); two EagleBurgmann HJ977/GN balanced end-face mechanical seals providing water confinement in the vessel while ensuring limited mechanical friction. In addition, this set of experiments has been performed under a $2.8$~bar pressure for optimal seal working conditions. 

\subsection{Mechanical measurements, Reynolds number}

In the first set of experiments, the mechanical physical quantities $f_1$ and $f_2$ (respectively the bottom and top impeller speeds) and $C_1$ and $C_2$ are measured using the monitor voltage outputs of the Yaskawa AC drives controlling the motors. The measurements of the second set of experiments rely on two Scaime MR12 torque sensors fixed to the impeller shafts, for more accurate measurements. In all experiments, the speed of the impellers is used to define the Reynolds number in the experiment, ${\rm Re}$:
\begin{equation}
	{\rm Re} = \frac{\pi (f_1 + f_2) R^2 }{\nu}
\end{equation}
The first set of experiments achieves Reynolds numbers in the range $800 \leq {\rm Re} \leq 185\,000$ and are therefore near the transition reported in~\cite{Ravelet2008}, whereas the Reynolds numbers of the second set of experiments are comprised between $100\,000$ and $400\,000$, well above the transition of~\cite{Ravelet2008}. For all speed imposed experiments, the rotation frequencies $f_1$ and $f_2$ of the impellers can be set from $1$~Hz to $12$~Hz, with a precision better than $0.2 \%$ in average whenever $f_1,\,f_2 \geq 2$~Hz. 

\begin{figure}
	\centering
	\includegraphics[width = 0.55 \textwidth]{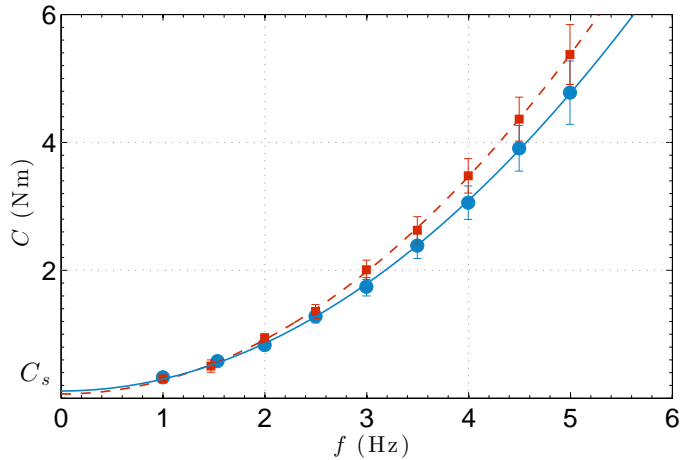}
	\caption{Determination of static friction torques $C_s$ in a calibration experiment. Red squares, top impeller torque, and blue circles, bottom impeller torque. Error bars represent the standard deviation of the torques at the speed plateaus. The static torques $C_{s1} = 0.11$~Nm and $C_{s2} = 0.067$~Nm are extrapolated from a quadratic fit of the torque at $f_1,~f_2 = 0$.}
	\label{fig:calib}
\end{figure}

For torque imposed experiments, an analogue 16-bit voltage source is used to impose the torques to the impellers. However, the torque actually transmitted to the impellers, $C_1$ and $C_2$, differs from this target torque due to torque sensor systematic offset and static friction of the mechanical seals.
 \ADD{The static friction is rather sensitive to the tuning of the seals and experimental conditions ; in addition, drifts of this friction have been reported between two consecutive weeks of experimental acquisitions}.
  Hence, all experiments require frequent calibrations (one a day), consisting in speed-imposed experiments performed at $f_1 = f_2$, with nine plateaus of the impeller speeds, the bottom impeller being started first (see \fref{fig:calib}). The mean torque values are then fitted by a quadratic law:
\begin{equation}
C = C_s + K_p f^2
\end{equation}
from which we identify $C_s$ to the static friction torque. The agreement between experimental data and this fit are good enough to eliminate both torque sensor bias and static friction, allowing a ``true" measure of the torques $C_1$ and $C_2$. Under such conditions, it is possible to assess the quality of the torque regulation along time: no deviation exceeding $0.5 \%$ of the average applied torque has been observed even in our longest experiments.  

\subsection{Velocimetry}

In addition to mechanical measurements, the second set of experiments includes fluid velocity measurement using a stereoscopic Particle Image Velocimetry (S-PIV) system provided by Dantec Dynamics. In this system, a Nd-YAG pulsed laser illuminates particles inside the cylinder in a vertical plane crossing the center of the experiment. Two digital cameras are filming the particles on two perpendicular faces of a square container surrounding the fluid vessel, to limit optical distortions due to air/water interfaces, producing a three-component velocity field ${\bf v} = (v_r, v_\phi, v_z)$ in a two-dimensional $(r,z)$ plane covering two azimuths $\phi_0$ and $\phi_0 + \pi$. Correlation calculations are performed on $32 \times 32$ and $16 \times 16$ pixels windows representing an area of $4.16 \times 4.16$~mm$^2$ and $2.08 \times 2.08$~mm$^2$, each time with an overlap between windows of 50$\%$. Our experiments typically yield $5000$ velocity maps on a $59 \times 63$ or a $113 \times 122$ grid. The time interval between two maps can be adjusted from $1/15$~s to $1~s$. 


\subsection{Symmetries and physical quantities associated}

The experimental set-up is characterized by two symmetries. The first one is the the rotational invariance along the $(Oz)$ axis, called axisymmetry, present for all rotation frequencies and torques. We will indeed assume that the addition of the blades on the discs has little impact on this symmetry: PIV measurements (see \fref{fig:pivmfs}) will confirm this assumption in the next sections. The other symmetry, shown in \fref{fig:setup}, is the $\mathcal{R}_\pi$ symmetry, which can be viewed as an upside-down flip of the experiment exchanging the two impellers. Obviously, a perfectly $\mathcal{R}_\pi$ symmetric flow requires $f_1 = f_2$ and $C_1 = C_2$. More generally, this symmetry allows a definition of four mechanical quantities, two being $\mathcal{R}_\pi$ symmetric, and two measuring the ``distance" to $\mathcal{R}_\pi$ symmetry:
\begin{equation}
	f = \frac{1}{2} \left ( f_1 + f_2 \right ),~C = \frac{1}{2} \left ( C_1 + C_2 \right ),~\theta = \frac{f_1 - f_2}{f_1 + f_2},~
	\gamma = \frac{C_1 - C_2}{C_1 + C_2} 
\end{equation}

Experiments performed at lower Reynolds numbers~\cite{Ravelet2008} at imposed speeds show that the von K\'arm\'an flow progressively breaks the symmetries of the mechanical setup: the instantaneous turbulent flow ${\bf v} (t)$ is neither $\mathcal{R}_\pi$ symmetric nor axisymmetric, even for symmetric forcing conditions. However, such symmetries are restored in a statistical sense~\cite{Frisch1995,Ravelet2004} and at large scales when we consider time-averages of a large number ($ \geq 600$) of PIV velocity maps (see fig.~\ref{fig:pivmfs}). These \emph{mean flows} are known to respect experimental and forcing symmetries.

To characterise the distance to $\mathcal{R}_\pi$ symmetry in the flow, we define the total kinetic angular momentum ${\bf I}(t)$ and its projection on the $z$ axis, $I(t)$:
\begin{equation}
I (t) = \frac{1}{\mathcal{V}} \int_{\mathcal{V}} \frac{r v_\phi}{2 \pi R^2 f} ~ r {\rm d} r {\rm d} z
\end{equation}
where $\mathcal{V}$ is the total volume of the fluid. It is simple to show that $\mathcal{R}_\pi$-symmetric flows produce distributions of $v_\phi$ cancelling $I$. This quantity is actually calculated on the two-dimensional grid of the PIV velocity field. We will once more consider that the values of $I$ obtained are representative of the whole volume of the cylinder, at least when $I$ is averaged on several samples, equivalent to an azimuthal average in our experiment.

\section{Evolution of the hysteresis cycle in speed control}
\label{hyst_evol_re}

\subsection{Qualitative aspect of the cycles and the flows}

\begin{figure}
	\includegraphics[width = \textwidth]{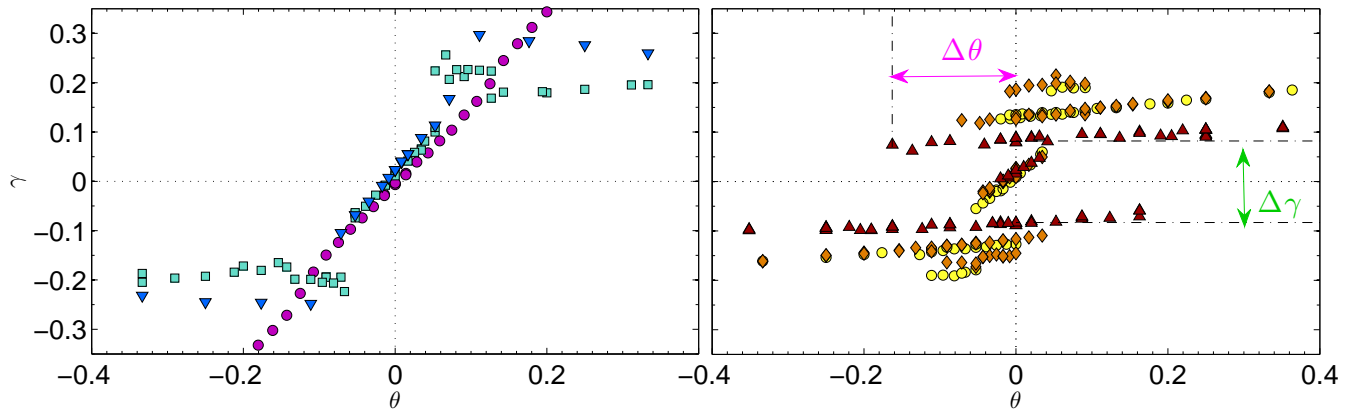}
	\caption{Evolution of the hysteresis cycle --- for impellers with 16 blades --- with ${\rm Re}$, from the laminar cases to hysteretic, turbulent flows. Left : Purple circles, ${\rm Re} = 800$; blue downwards triangles, ${\rm Re} = 2900$; teal squares, ${\rm Re} = 4900$. Right: yellow circles, ${\rm Re} = 9500$ (symmetrised); diamonds ${\rm Re} = 15\,500$; upwards triangles, ${\rm Re} = 185\,000$.}	
	\label{fig:open_cycle}
\end{figure}

In this section, we will study the global, mechanical response to an asymmetric field in \emph{speed control}, making use of the first set of experiments. This study follows the direct visualisations and the Doppler velocity measurements performed in~\cite{Ravelet2008} to describe the transition to turbulence in the von K\'arm\'an flow. The experimental results, displayed on~\fref{fig:open_cycle}, show for the lower Reynolds numbers (${\rm Re} = 800$ corresponding to chaotic, yet not turbulent, velocity fields), a continuous response to an asymmetric field. For slightly higher Reynolds numbers (${\rm Re = 2900}$), a slight discontinuity of the average response $\gamma$ can be observed for non-zero asymmetries ($\theta \approx \pm 0.09$), whereas the velocity spectrum of corresponding experiments in~\cite{Ravelet2008} exhibits stronger small-scale fluctuations, with a beginning of power-law decay corresponding to the inertial range. The discontinuities appear more clearly for ${\rm Re} = 4900$, and are respectively visible for $\theta = \pm -0.07$ and $\theta = \pm 0.13$. In this picture, the mechanical response displays no hysteresis, the velocity fluctuations have saturated (see~\cite{Ravelet2008}) and an inertial range is well visible in the velocity spectra: \ADD{turbulence in the flow can therefore be considered developed.}

For ${\rm Re} \geq 10\,000$ and $\theta = 0$, a new statistical state called \emph{bifurcated} state has been evidenced (see figure~7 in~\cite{Ravelet2008} for details), showing that bifurcations can be (at least statistically) observed in a turbulent flow. \Fref{fig:open_cycle} (right) reveals, for ${\rm Re} \approx 9500$ and ${\rm Re} \approx 15\,500$, hysteretic asymmetry responses with five superposed branches. The main difference between these two very similar cycles is that the four non-central branches are smaller for ${\rm Re} = 9500$ with two small outermost branches not crossing $\theta = 0$, while the cycle at ${\rm Re} = 15\,500$ shows five coexisting branches for $\theta = 0$. These branches are already present at ${\rm Re} = 4900$ but have extended to $\theta = 0$ and beyond with an increase of the Reynolds number, leading to a strong hysteresis. The discontinuity of the normalised torque (visible in~\cite{Ravelet2008}, on figure~7) is therefore already present at lower Reynolds numbers, and results from a continuous evolution where the branches progressively superpose each other: additional data (not presented in this article) show that the four external branches are already superposed for ${\rm Re} = 6700$. Eventually, two of the five branches present at ${\rm Re \approx 15\,500}$ vanish for the highest Reynolds numbers achievable in our experiment, leading to a ``typical" hysteresis cycle for ${\rm Re} \geq 100\, 000$, as visible in \fref{fig:open_cycle}. It is extremely similar to the cycle with eight blade impellers (see \fref{fig:gitv} for a ${\rm Re} \approx 250\,000$ cycle): $(b_1)$ and $(b_2)$, called bifurcated branches, surround a central, symmetrical branch called $(s)$. 

This $(s)$ branch is only accessible starting both impellers at the same time and exhibits --- in average --- a steady and $\mathcal{R}_\pi$ symmetric turbulent flow. Each impeller generates a recirculation cell covering half of the total fluid volume and imposes a global rotation of the aforementioned volume (see \fref{fig:pivmfs}). At altitude $z = 0$, a \emph{shear layer} can be defined in the average, steady flow (see~\cite{Cortet2011} for details). In this branch, slightly asymmetric experiments display a finite lifetime before transiting abruptly (and irreversibly) to another steady state corresponding to the $(b_1)$ (or $(b_2)$) branch. The lifetime of these weakly asymmetric steady states diverges when the asymmetry parameter, $\theta$, tends to zero~\cite{Ravelet2005}: the $(s)$ branch is thus called marginally stable.

In contrast, $(b_1)$ and $(b_2)$ flows are strikingly different, breaking in average the $\mathcal{R}_\pi$ symmetry even when $\theta = 0$.  For $\theta = 0$, these states can be selected applying transient asymmetries, starting for example one impeller before the other. A dramatic raise of the mean torque is also observed these two branches, up to $3.4$ times the torque level of the $(s)$ branch. The $(b_2)$ ---~selected for example by starting the top impeller first~---state consists of one recirculation cell where the fluid at the centre of the cylinder is pumped by the top impeller for all altitudes $z$. In addition, the top impeller imposes a global rotation of all the fluid (see \fref{fig:pivmfs}). Obviously, the mean flow selected when the bottom impeller is started first is the image by $\mathcal{R}_\pi$ symmetry of such a flow. The hysteresis of these branches reflects the fact that the top (respectively bottom) impeller is able to pump all the fluid even though the bottom (respectively top) impeller rotation speed is higher. In these configurations, the impeller that is pumping the fluid always provides the largest torque, despite a lower rotation rate. This can be seen in figure 3, noting that the bifurcated branch with $\gamma>0$ ($\gamma<0$) extends far in the $\theta<0$ ($\theta>0$) domain. 

\ADD{The impeller blade curvature plays here a key role in the shape of the turbulent asymmetry response. A thorough study by Ravelet~\cite{Ravelet2005, Ravelet2004} has reported that this response is continuous for small or null blade curvature, discontinuous with no hysteresis for intermediate curvature, and eventually hysteretic for high curvature --~the impellers of the article falling into the latter category~-- in the $(-)$ rotation sense. This study is consistent with the existing literature where only straight blades are used (as seen on the sketches and pictures of~\cite{Labbe1996,Voth2002,Titon2003,Mordant2004,Ouelette2006}), or curved blades in the reverse, $(+)$ rotation sense~(see \cite{Torre2007,Saint-Michel2013a}). Curved blades imply a larger toroidal flow~\cite{Ravelet2005} which tends to favour hysteresis: to illustrate this, Ravelet used vertical baffles in the cylinder slowing down the toroidal flow, significantly reducing the hysteresis present in the experiment~\cite{Ravelet2004}.}

\subsection{A more quantitative approach}
A finer description of the evolution of the hysteresis loop can be formulated studying the height $\Delta \gamma$ and the $\Delta \theta$ of all cycles. These parameters can be mapped to their corresponding counterparts in the hysteresis cycles of ferromagnetic materials. Under a critical temperature $T_c$, such materials exhibit spontaneous magnetization in the absence of any external magnetic field: our cycle height $\Delta \gamma$ can therefore be seen as the ``spontaneous magnetisation" of our flow. Similarly, ferromagnetic spontaneous magnetisation cancels for a particular magnetic field strength referred as the coercive field: our cycle width $\Delta \theta$ is therefore its von K\'arm\'an counterpart. 

 Similarities with magnetic materials have already been evidenced in the other rotation direction, $(+)$ (see~\cite{Cortet2011} for details). The complex nature of some cycles requires a careful definition of the cycle height, therefore, two definitions of this height have been considered. The first definition simply examines if the bifurcated branches (identified as the branches extending to $\theta = \pm 1$) extend to $\theta =0$. If they do, the height is measured as the distance between the two branches. In the other case, $\Delta \gamma$ is zero. The second definition of the height is basically the same, except that the $(b_1)$ and $(b_2)$ branches are linearly extended to $\theta = 0$. Both definitions are identical in strongly hysteretic cycles, hence, for ${\rm Re \geq 7800}$.

\begin{figure}
	\includegraphics[width = \textwidth]{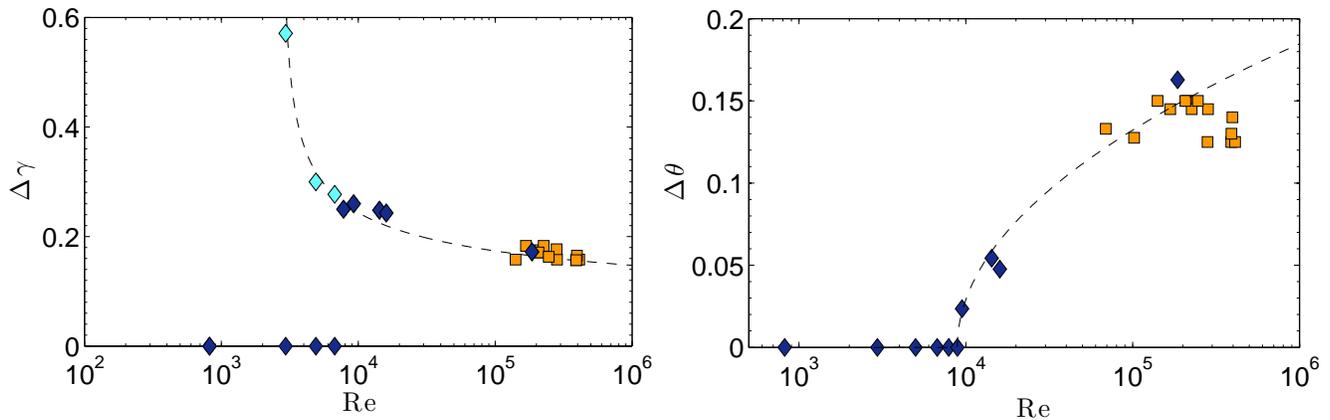}
	\caption{Left: height of the hysteresis cycle $\Delta \gamma$ defined as the distance (in terms of $\gamma$) between branches $(b_1)$ and $(b_2)$ for $\theta = 0$. Cyan diamonds, impeller with 16 blades, where the outermost branches have been prolonged to $\theta = 0$. Navy diamonds, same set of data, without prolongation. Orange squares, experiments with 8 blade impellers. Right: width of the hysteresis cycle, defined as the distance (in terms of $\theta$) between the tips of the $(b_1)$ and $(b_2)$ branches. Same symbols as before. The dashed lines represent adjustments using power laws of $\left | 1 / \log({\rm Re})  -  1 / \log({\rm Re}_c) \right |$ (see text).}	
	\label{fig:param_cycle}
\end{figure}

Quite surprisingly, \fref{fig:param_cycle} reveals a  torque asymmetry $\Delta \gamma$ bifurcating from a very large value at ${\rm Re}\sim 2900$, whereas the coercive field $\Delta \theta$ bifurcates from zero at  ${\rm Re} \sim 8900$, both quantities following roughly a power-law pas the transition as: 
\begin{equation}
	\left | \frac{1}{\log({\rm Re})} - \frac{1}{\log({\rm Re}_c)} \right |,
\end{equation}
following a definition of the ``temperature" of turbulent flows suggested by Castaing~\cite{Castaing1996}. The experimental data roughly respect the following theoretical power-laws:
\begin{eqnarray}
	\Delta \gamma & \propto    K - & \left   | \frac{1}{\log({\rm Re})} - \frac{1}{\log({\rm Re}_1)} \right |^\alpha,~ \alpha = \frac{1}{6},~ {\rm Re}_1 = 2900 \\
	\Delta \theta & \propto &  \left | \frac{1}{\log({\rm Re})} - \frac{1}{\log({\rm Re}_2)} \right |^\beta,~ \beta = \frac{2}{3},~{\rm Re}_2 = 8900
\end{eqnarray}

The presence of power laws suggests the presence of a critical transition for two Reynolds numbers, one of them roughly corresponding to the transition to turbulence,
 \ADD{associated to the emergence of a noticeable region of $k^{-5/3}$ spectrum of the velocity spectrum on previous Laser Doppler Velocimetry measurements~\cite{Ravelet2005}}
 , the other one being observed for even higher Reynolds numbers. There is at the present time no explanation for the value of the exponents $\alpha$ and $\beta$. 

\ADD{It is also unknown whether these power lows survive until $\mathrm{Re} \rightarrow \infty$, or if $\Delta \gamma$ and $\Delta \theta$ reach asymptotic, finite values. In critical phenomena, such power laws are only valid in the vicinity of the critical temperature, for the exact relations --~when they can be derived~-- between relevant physical quantities and temperature can be far more complex (see Onsager's exact solution of the two-dimensional Ising model for instance). This finite asymptotic regime, if it exists, seems currently beyond the reach of our experimental setups.}

\ADD{Nevertheless, since the power-laws previously described involve Reynolds dependences through $1 / \log {\rm Re}$ terms, we will consider in the rest of the article that far from the critical Reynolds numbers (for ${\rm Re} \geq 150\,000$), the cycle characteristics are only weakly dependent on the values of \emph{${\rm Re}$} and the number of blades: more specifically, doubling the impeller speed should not result in more than a $10 \%$ variation of the cycle parameters (which can be verified in~\fref{fig:param_cycle}). In addition, the global shape of the hysteresis cycle remains unchanged for such Reynolds numbers. These assertions have been confirmed by experiments performed in water~\cite{Saint-Michel2013a} and in liquid helium at larger ($10^8$) Reynolds numbers~\cite{Saint-Michel2014}.}

\section{Influence of the forcing nature at high Reynolds numbers}
\label{two_controls}

This section is based on the second set of experiments. It will focus on the influence of the forcing nature on the ---~statistically~--- steady states reached when an asymmetric field is imposed in the experiment, both in speed and torque controls. It is indeed always possible to report the average values of the response (respectively $\gamma$ in speed control and $\theta$ in torque control) to an imposed asymmetry (respectively $\theta$ in speed control, or $\gamma$ in torque control) in the same $\gamma,~\theta$ diagram. PIV and mechanical measurements will be used to complete the characterisation of the steady states sketched in previous work~\cite{Saint-Michel2013} using only mechanical measurements.

\subsection{Speed control and torque control}

\begin{figure}
	\centering
	\includegraphics[width = \textwidth]{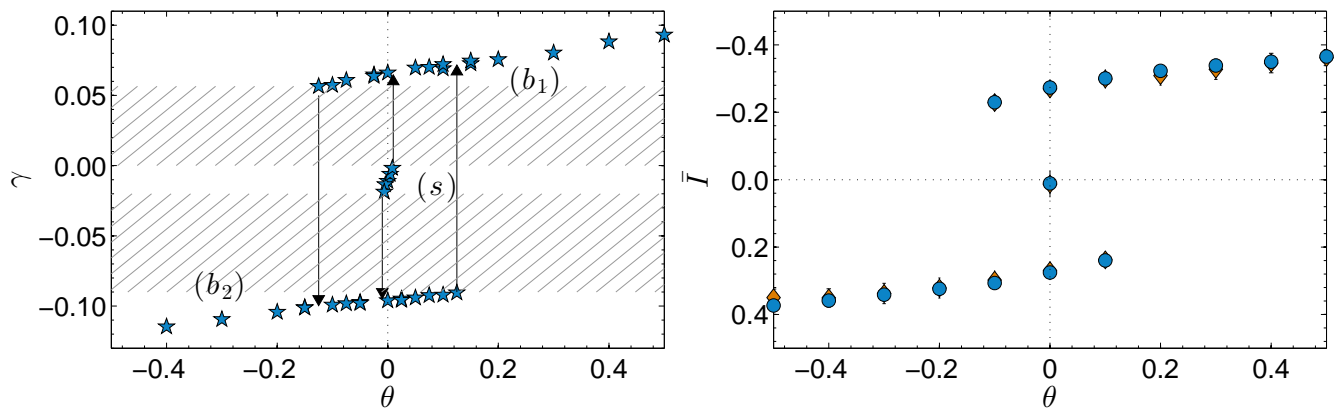}
	\caption{Left: mean reduced torque asymmetry $\gamma$ ---~teal stars~--- as a function of $\theta$ for all the steady states accessible in speed control, for experiments performed at $f = 4$~Hz (${\rm Re} \approx 250\,000$). The center branch, $(s)$, is marginally stable, and the external branches $(b_1)$ and $(b_2)$ are hysteretic. The arrows indicate the various transitions observed between steady states: the center branch cannot be accessed from another branch, and is therefore \emph{marginally stable}. The hatched region represents a $\gamma$ range never explored by the flow. Right: reduced mean angular momentum $\bar{I}$ for the three branches in speed control. Blue circles, experiments at a rotation speed of $f = 4$~Hz experiments, and orange diamonds, experiments at $f = 5$~Hz. The y axis has been inverted to emphasize the similarities between $I$ and $\gamma$.}
	\label{fig:gitv}
\end{figure}

\subsubsection{Comparison between PIV and global measures in speed control}

As already stated in the previous section, speed-control turbulent flows are statistically steady and display a large hysteresis cycle in the $(\gamma, \theta)$ plane, composed of three continuous branches separated by a forbidden zone ---~a range of $\gamma$ values~--- where no steady state is reached. This zone is represented by a hatched region in \fref{fig:gitv}. It is also possible to visualise the hysteresis cycle using the mean angular momentum $\bar{I}$ as a response to the asymmetric field $\theta$. \Fref{fig:gitv} highlights the similarities between such cycles: $\bar{I}$ and $\gamma$ seeming almost proportional to each other, the hysteresis cycle plotted with $\bar{I}$ being only weakly dependent on the impeller rotation frequency $f$.

\begin{figure}
	\centering
	\includegraphics[width = 0.6 \textwidth]{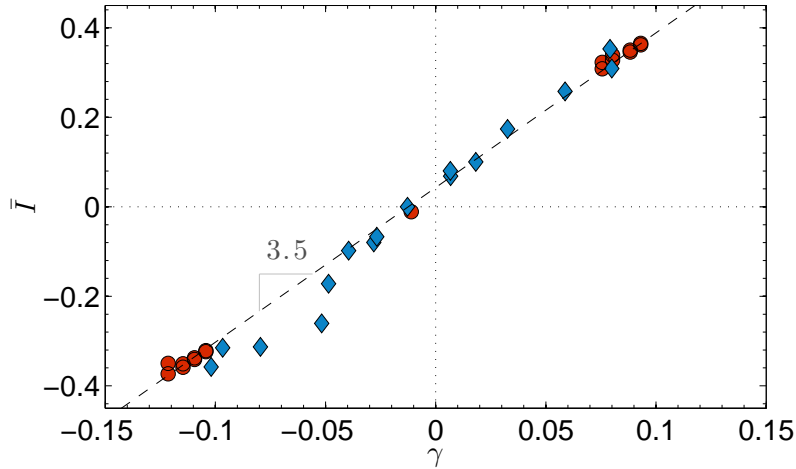}
	\caption{Mean reduced angular momentum $\bar{I}$ as a function of the torque asymmetry $\gamma$ in speed control. Red circles experiments are performed in the rotation direction described in \fref{fig:setup}, and leaves gaps between the central and extremal values. For this particular rotation direction, the y-axis is reversed.
	 Values of $\bar{I}$ in the gap can be evaluated by computing the same relation in the other rotation direction (teal diamonds) for which the y-axis is not reversed. Both senses exhibit a very similar, affine relation.}
	\label{fig:gammaI}
\end{figure}

We can evaluate such a similarity by plotting $\bar{I}$ as a function of $\gamma$. \Fref{fig:gammaI} confirms an affine relation between these quantities: $\bar{I} = a \gamma + b$ , with $a = 3.6 \pm 0.1$ and $b = 0.037 \pm 0.01$. The offset observed at ${\bar I} = 0$ is interpreted as a residual mechanical asymmetry which is not accounted for in the calibrations. This relation is actually valid for both directions of rotation of the discs at high Reynolds numbers. This simple link between PIV and mechanical measurements is very useful considering the time required for PIV data processing compared to mechanical measurements, and the limited duration of PIV experiments. This link also indicates that the results of~\cite{Cortet2011} can be extended to torque measurements.

\subsubsection{Corresponding torque experiments}
\label{torqueexp}

\begin{figure}
	\centering
	\includegraphics[width = \textwidth]{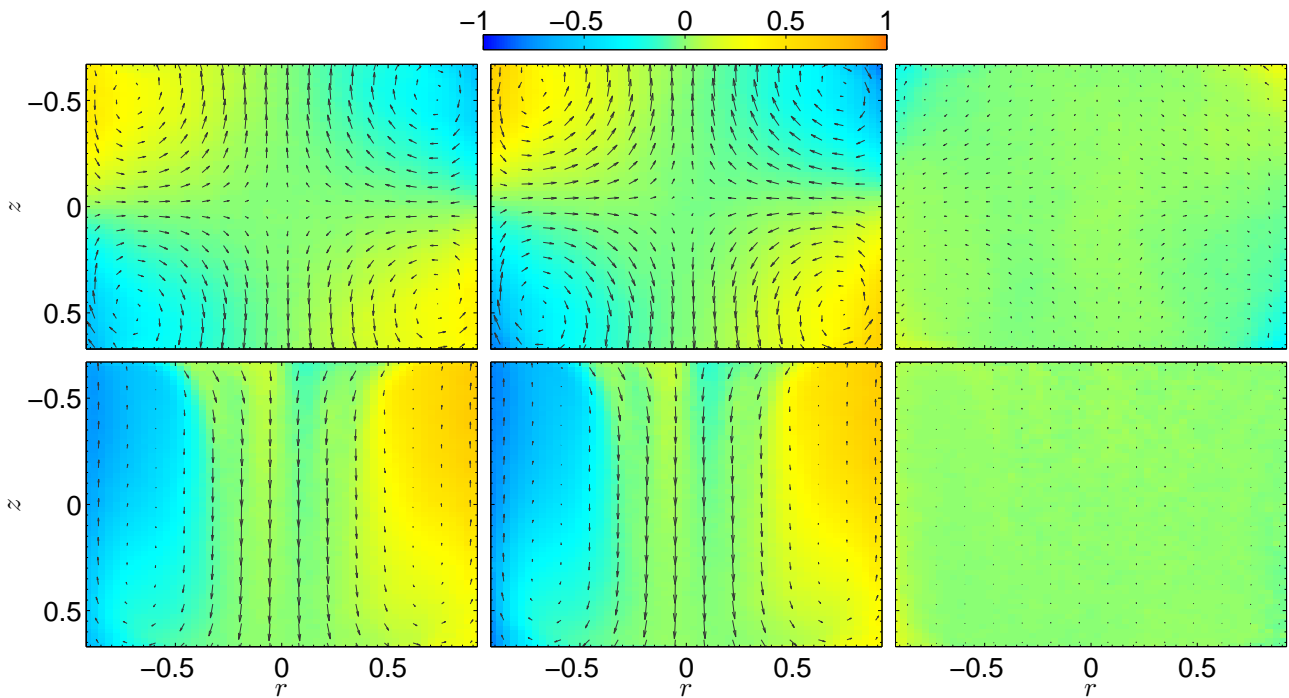}
	\caption{Mean turbulent flows ${\bf v} / (2 \pi R f)$ produced by stereoscopic PIV for a turbulent experiment: ${\rm Re} \approx 250\,000$. From left to right, mean flow in torque control, mean flow in speed control, and difference between the mean flows. Top: flows in the symmetric branch $(s)$ for $\theta \approx 0$, and bottom, flows in the bifurcated branch $(b_1)$ for $\theta = 0$. Arrows, poloidal velocity $(v_r, v_z)$, and colours, azimuthal velocity $v_\phi$. The arrow scale used for the $(s)$ flow is different from the one used for $(b_1)$ flows for clarity. The frequency used to normalise ${\bf v}$ in torque control is the mean impellers speed $\bar{f}$ during the experiment. Sign conventions imply that positive values of $v_\phi$ always represent a net movement towards the observer. The measurements made at azimuth $\phi_0 + \pi$ are represented in the $r < 0$ part of the velocity map.}
	\label{fig:pivmfs}
\end{figure}

In torque control experiments, the lower and upper torques $C_1$ and $C_2$ (and therefore $\gamma$) are imposed while the speed of the impellers $f_1$ and $f_2$ is free to evolve. The actual values of $C_1$ and $C_2$ (and, therefore, of $\gamma$) are unknown before a calibration assessing the friction torques and the sensor bias is performed. First, torque control experiments have been conducted for values of $\gamma$ corresponding to the speed control steady states. For two values of $\gamma$, corresponding to  $\theta = 0$ $(s)$ and $\theta = 0$ $(b_1)$, the velocity maps are extremely similar (see \fref{fig:pivmfs}): the minor residual differences are attributed to uncertainties on the exact values of $\gamma$ and $\theta$ for both controls.

The $\theta, \gamma$ steady states produced in torque control (see \fref{fig:gthall} for a superposition) are very well superposed with their speed control counterpart, but the mean impeller speed of the steady $(b_1)$ and $(b_2)$ states is lower than in the $(s)$ branch. However, since the Reynolds number dependence of the shape of the $\gamma, \theta$ diagram is slow for this Reynolds number range, we will suppose that, out of the ``forbidden zone", both controls yield steady states spanning the same ``master" $(\gamma, \theta)$ curve. Dimensional analysis based on the difference of mean torques $C$ observed in speed control provides an estimation of this speed ratio, yielding:
\begin{equation}
\frac{f_{(s)}}{f_{(b)}} = \sqrt{ \frac{C_{(b)}}{C_{(s)}} } \approx \sqrt{3.4} = 1.84
\end{equation}
The experimental ratio $f_{(s)} / f_{(b)}$ for torque control experiments is $1.86$, consistent with our predictions.

\subsubsection{Forbidden zone experiments}

\begin{figure}
	\centering
	\includegraphics[width = \textwidth]{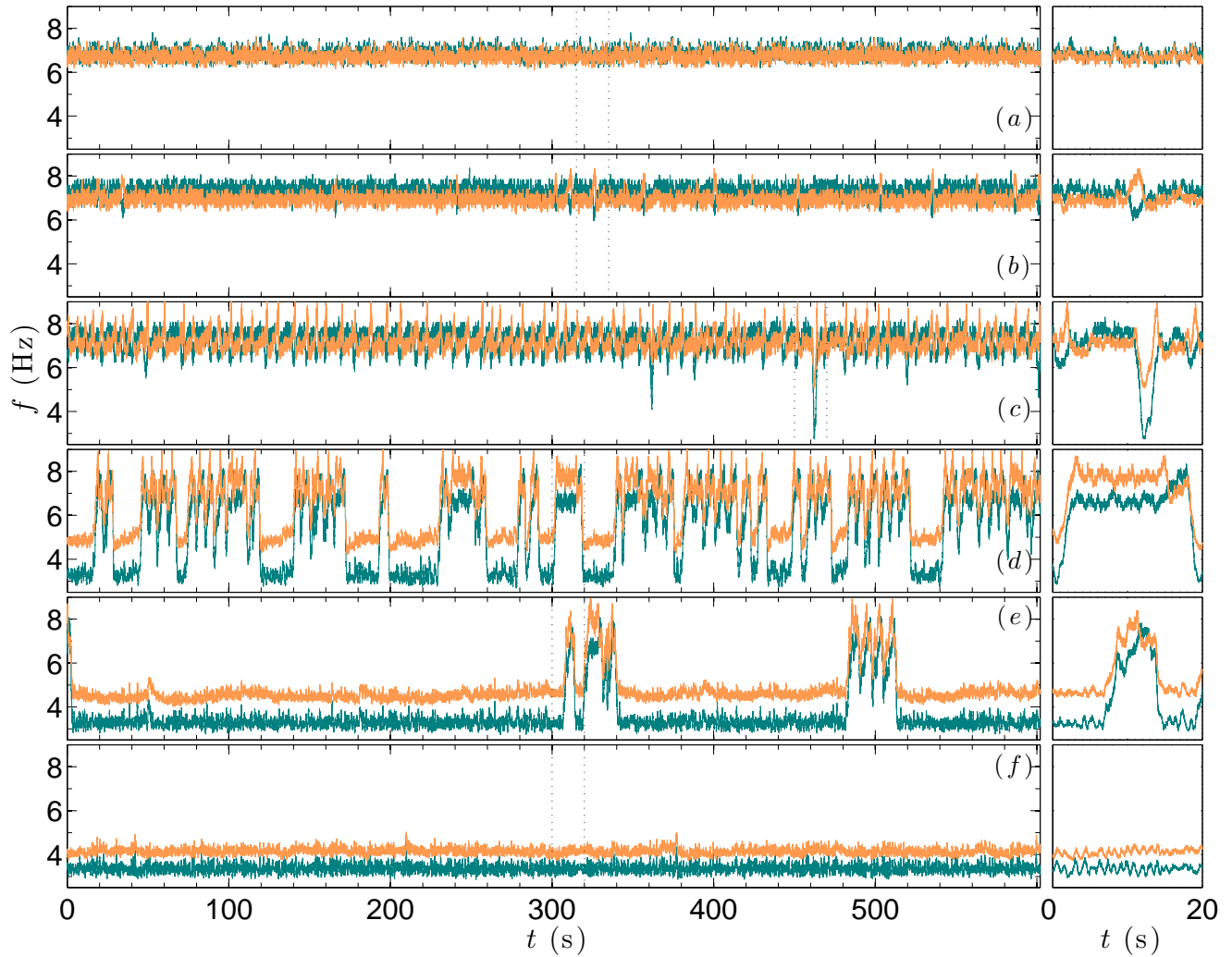}
	\caption{Time series of torque control experiments for various values of $\gamma$. Dark teal line, bottom impeller. Orange line, top impeller. Left panels, $600$~s of experimental data. Right panels, $20$~s zoom on characteristic events (when such events occur) delimited by the dotted line on the left panels. $(a)$, steady experiment corresponding to $(s)$ for $\gamma = -6.1\cdot 10^{-3}$; $(b)$, small excursions observed for $\gamma = 0.014$; $(c)$, threshold of big excursions for $\gamma = 0.026$; $(d)$, multi-stable experiment for $\gamma = 0.045$; $(e)$, rare events for $\gamma = 0.054$; $(f)$, steady $(b)$ experiment for $\gamma = 0.059$}
	\label{fig:tempseries}
\end{figure}

The main difference between torque control and speed control resides in the ability to choose values of $\gamma$ corresponding to the speed control ``forbidden" zone. Starting from a $\mathcal{R}_\pi$ symmetric flow in the $(s)$ branch, we can progressively increase the value of $\gamma$. First, the flow remains steady, with a slightly increased $f_1$ and decreased $f_2$ until $\gamma = 3.5 \cdot 10^{-3}$ (see \fref{fig:tempseries}$a$). Then, small jumps of the speed of the impellers are observed, as depicted in \fref{fig:tempseries}$b$: on each of these events, the faster impeller suddenly slows down. A local maximum of the
slower impeller speed follows two impeller revolutions later, after which the impellers return to their initial speeds. These jumps have a characteristic duration of $3$~s (20 impeller rotations), and thus break the statistical time invariance of the flow. We can interpret  \fref{fig:tempseries}$b$ time series as a general system remaining most of the time in an attracting state corresponding to $(s)$ and escaping from time to time this attractor during excursions (named after the similar excursions of the magnetic field observed in the von K\'arm\'an Sodium experiment~\cite{Berhanu2007}) towards another unknown attracting state named $(i_1)$, or \emph{intermediate} attracting state. Increasing the value of $\gamma$ increases the number of excursions towards $(i_1)$,until the impeller speeds $f_1$ and $f_2$ are nearly periodic.

 For $\gamma \geq 0.026$, a new type of event is visible, as seen in \fref{fig:tempseries}$c$: both impellers drastically slow down, reach a minimum speed at the same time, and speed up again, the least-forced impeller reaching afterwards a local speed maximum. Still increasing $\gamma$, our experiments reveal a multi-stability between multiple attracting states $(s)$, $(i_1)$, and a slow quasi-steady state corresponding to $(b_1)$. This regime is clearly visible in \fref{fig:tempseries}$d$ where we can observe an $(i_1)$ event exceeding 13~$s$ (or 75 impeller revolutions). For larger values of $\gamma$ (\fref{fig:tempseries}$e$), a ``rare event" regime is observed, exhibiting very long periods of $(b_1)$ (up to $1-2$ hours) interspersed with sudden accelerations of both impellers towards the $(i_1)$ and $(s)$ states of relatively short duration, generally lasting from $10$ to $40$~seconds. This variable duration suggests a different nature than the excursions from $(s)$ to $(b_1)$ which always lasted the same amount of time. Eventually, for $\gamma$ values corresponding to the speed control $(b_1)$ steady states, steadiness is restored, as already stated on section~\ref{torqueexp}.

\subsection{The quasi-steady states}

\subsubsection{PIV characterisation}

\begin{figure}
	\centering
	\includegraphics[width = 0.9 \textwidth]{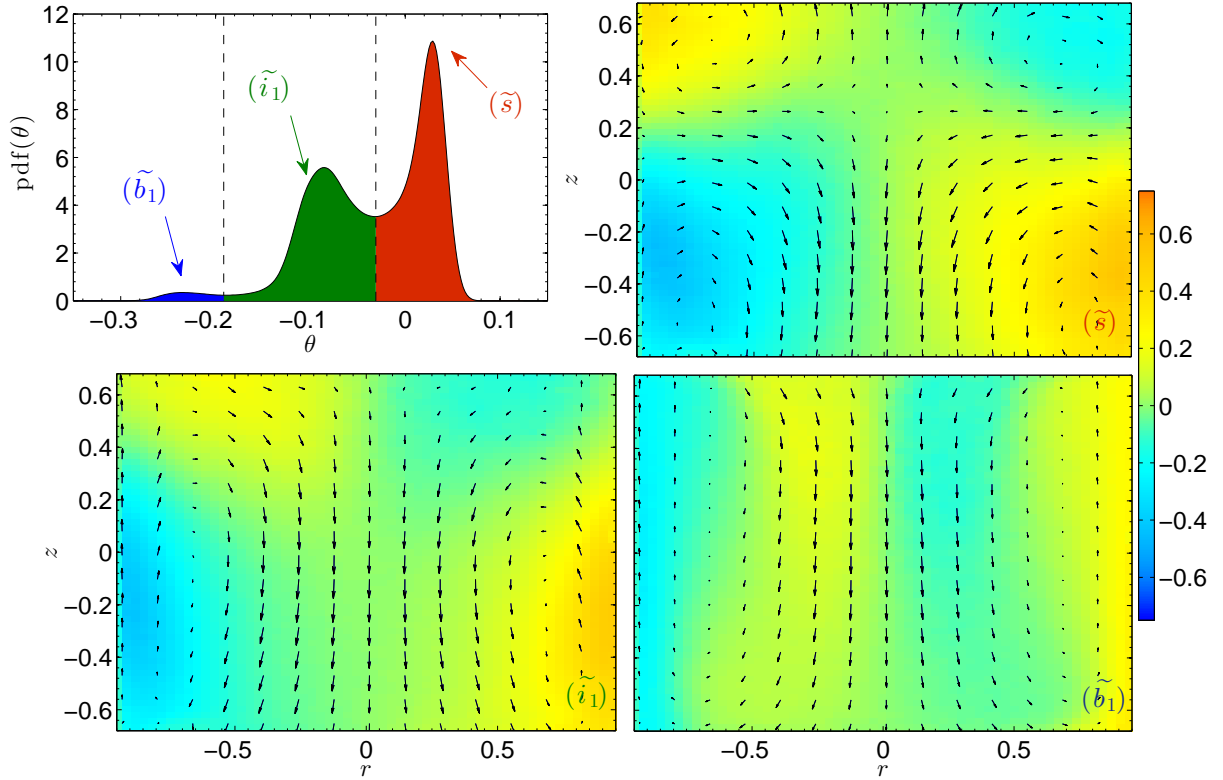}
	\caption{Top-left: p.d.f. of the filtered signal of $\theta(t)$ using a kernel-smoothing density estimate, for a $\gamma = 0.028$ experiment of average Reynolds number ${\rm Re} \approx 400\,000$. The local minima define boundaries of $\theta$ ranges identified with the quasi-steady states $(b_1)$, $(i_1)$ and $(s)$, used for conditional averaging of velocity fields. Top-right: conditional average flow corresponding to the $(s)$ quasi-steady state. Bottom-left: conditional average velocity field corresponding to $(i_1)$. Bottom right, conditional average $(b_1)$ velocity.	Following the decision made in \fref{fig:pivmfs}, the scale used for the arrows corresponding to the $(b_1)$ velocity map is smaller than those used for both $(i_1)$ and $(s)$. The conventions used for the sign of $r$ and the choice of colours for $v_\phi$ are explained in \fref{fig:pivmfs}.}
	\label{fig:pdf_mfcond}
\end{figure}

The velocity maps corresponding to the steady $(s)$ and $(b)$ states have already been described in \fref{fig:pivmfs}. However, the nature of the $(i)$ mean velocity field is unknown, and the \emph{quasi}-steady mean $(s)$ and $(b_1)$ velocity fields might differ from their steady counterparts. Hence, conditional mean fields have been computed using post-synchronised PIV and mechanical measures, for an experiment with $\gamma = 0.028$ (between \fref{fig:tempseries}$c$ and \fref{fig:tempseries}$d$). The averaging condition is based on the $1$~Hz filtered signal of the reduced speed asymmetry, $\theta(t)$, using a kernel-smoothing density estimation. In such regions, the p.d.f (probability density function) of $\theta(t)$ displays multiple peaks separated by local minima defining the $\theta$ boundaries of the $(s)$, $(i_1)$ and $(b_1)$ used for conditional averaging (see \fref{fig:pdf_mfcond} for an example). Surprisingly, while $\gamma$ is manifestly positive, the two peaks corresponding to $(i_1)$ are both located in $\theta \leq 0$ regions ---~ $(i_1)$ peaking around $\theta = -0.085$) and $(b_1)$ near $\theta = -0.235$)~--- and thus tend to decrease the average, experimental value of $\theta$ in the forbidden region. Such diminution of the average $\theta$ feels however natural if we consider our transition from pure $(s)$ to pure $(b_1)$ in torque control to be a continuous process: connecting the branches of \fref{fig:gitv} in such a waynecessarily requires to decrease $\theta$ while increasing $\gamma$.

\begin{figure}
	\centering
	\includegraphics[width = \textwidth]{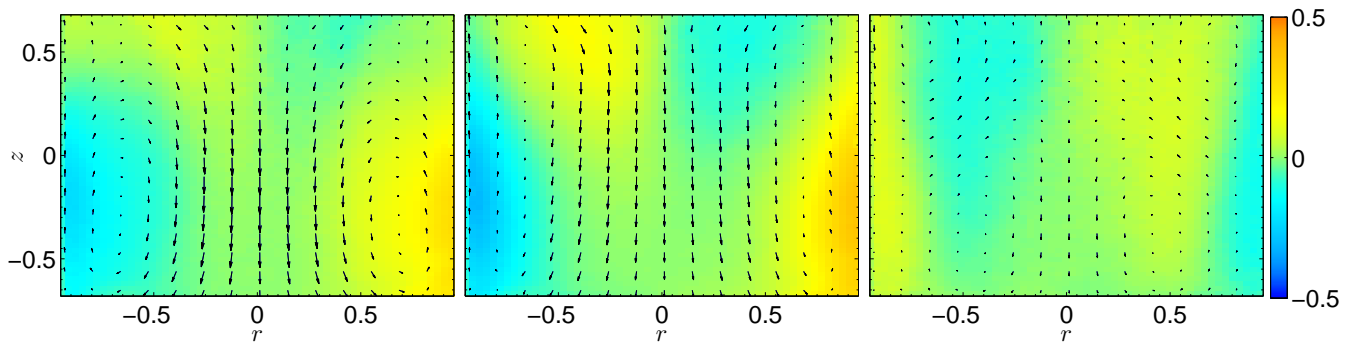}
	\caption{Comparison between the synthetic intermediate mean velocity field ${\bf v}_i^s$ (left) defined as a linear combination of the conditional average fields ${\bf v}_s$ and ${\bf v}_b$ of a multi-stable experiment minimising the distance with the intermediate velocity field ${\bf v}_i$, and the actual intermediate average velocity ${\bf v}_i$ (center). The right panel plots ${\bf v}_i^s - {\bf v}_i$ and shows a non-negligible residual difference, in particular for $v_\phi$. Same graphic conventions as in \fref{fig:pivmfs} and \fref{fig:pdf_mfcond}.}
	\label{fig:compareisb}
\end{figure}

Following these conventions, it is possible to extract the three conditional flows corresponding to the quasi-steady states. The $(s)$ state (\fref{fig:pdf_mfcond}, top right) presents two circulation cells but breaks $\mathcal{R}_\pi$ symmetry with a larger lower circulation cell: this velocity map is very similar to the steady states observed for small $\theta$ in the $(+)$ rotation sense of~\cite{Cortet2011}. The $(b_1)$ conditional flow bears similarities to the $(b_1)$ steady flow described in \fref{fig:pivmfs}: the bottom impeller generates a single pumping cell, drawing the fluid at low $r$ to eject it at larger values of $r$. However, the impeller is unable to impose a global rotation of the whole fluid volume: for low $r$ values, the fluid seems to rotate in the inverse direction corresponding to the least-forced impeller. The intermediate $(i_1)$ conditional average flow is peculiar: whereas all the fluid is pumped --- for $\gamma > 0$ by the lower impeller, the azimuthal flow exhibits two rotation cells separated by a V shaped shear layer. It is therefore tempting to consider such a flow to be a simple mix between the $(s)$ and the $(b_1)$ conditional flows, $(i_1)$ only resulting from an accumulation of transitions between $(s)$ and $(b_1)$. Hence, \fref{fig:compareisb} compares the PIV average velocity field ${\bf v}_i$ from the intermediate state to the synthetic field ${\bf v}_i^{s}$ computed from a linear combination of ${\bf v}_s$ and ${\bf v}_b$ and minimising the distance $\mathcal{D}$ to the experimental conditional mean field:
\begin{equation}
\mathcal{D} = \left | {\bf v}_i^s - {\bf v}_s \right | = {\rm min}_{(\alpha, \beta)} \int_{\mathcal{V}} ( \alpha {\bf v}_s + \beta {\bf v}_b - {\bf v}_s ) \cdot ( \alpha {\bf v}_s + \beta {\bf v}_b - {\bf v}_s )~ r {\rm d} r {\rm d} z~
\end{equation}

\Fref{fig:compareisb} shows that the synthetic field ${\bf v}_i^s$ can reproduce some of the features of the intermediate velocity field, but not all of them: a significant difference on the shape of the azimuthal velocity (especially at high $r$ and $z$) can be observed. Minor differences in the poloidal velocity can also be found, the synthetic flow producing a higher vertical average flow than its ``true" counterpart. Thus, the $(i)$ quasi-steady state is new and distinct from $(s)$ and $(b)$.

\subsubsection{Position on the hysteresis cycle}

\begin{figure}
	\centering
	\includegraphics[width = 0.75 \textwidth]{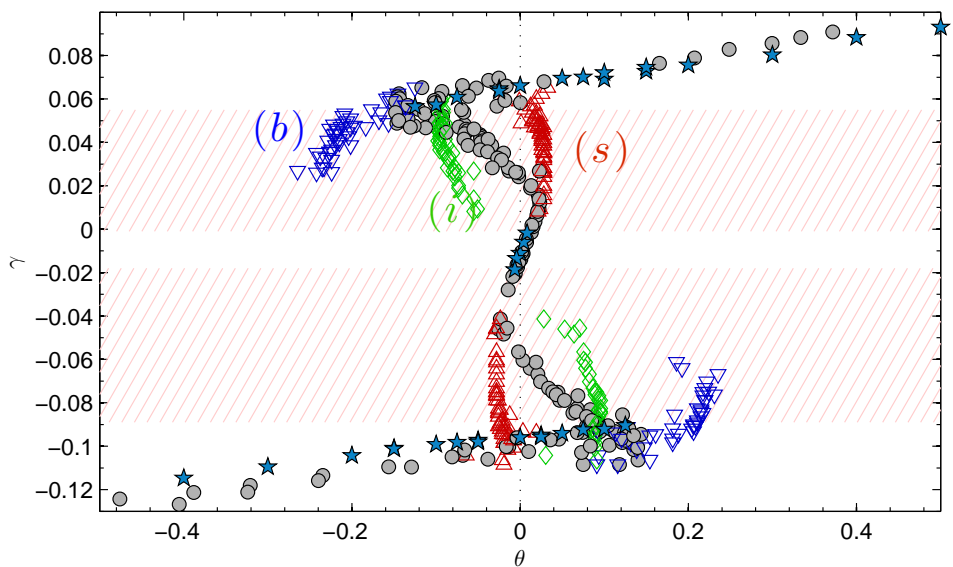}
	\caption{Full diagram of the steady and unsteady experiments in speed and torque control on the $\theta, \gamma$ plane. Teal stars, reminder of the speed control hysteresis cycle. Grey circles, mean value of $\theta$ for all torque control experiments. Red (pointing upwards) triangles, local maximum of $\theta$ corresponding to $(s)$ in unsteady torque control experiments. Green diamonds, intermediate $(i)$ local maximum of $\theta$. Blue (pointing downwards) triangles, $(b)$ local maxima. The hatched areas represent the speed control forbidden zone.}
	\label{fig:gthall}
\end{figure}

The local maxima of the distribution of $\theta$ corresponding to $(s)$, $(i)$ and $(b)$ for unsteady experiments can be superposed to the $\gamma, \theta$ diagram of \fref{fig:gitv} to evaluate the relative positions of the steady $(s)$ and $(b)$ branches, their unsteady counterpart, and the intermediate branches. \Fref{fig:gthall} shows a synthetic plot of our dynamics: the forbidden zone in speed control is associated with negative slopes of the average value of $\theta$ in torque control, which is itself associated with the emergence of the multiple peaks of the p.d.f of $\theta$. In addition, the $(b)$ and $(s)$ unsteady branches (respectively blue and red triangles) behave like extensions of the steady $(b)$ and $(s)$ branches, validating our interpretation of these maxima. Whereas $\theta_b$, the abscissa of the $(b)$ local maximum, seems to restore positive slopes $\partial \theta_b / \partial \gamma$, the corresponding values for $(i)$ and (to a lesser extent) $(s)$ still exhibit negative slopes. Finally, the location of the intermediate branches is still unclear and seems to span from an unknown location near the steady $(s)$ center branch to the border of the $(b)$ steady branch.

\subsection{Dynamics between quasi-steady states}

\subsubsection{Excursion and transition patterns}
\begin{figure}
	\centering
	\includegraphics[width  = 0.95 \textwidth]{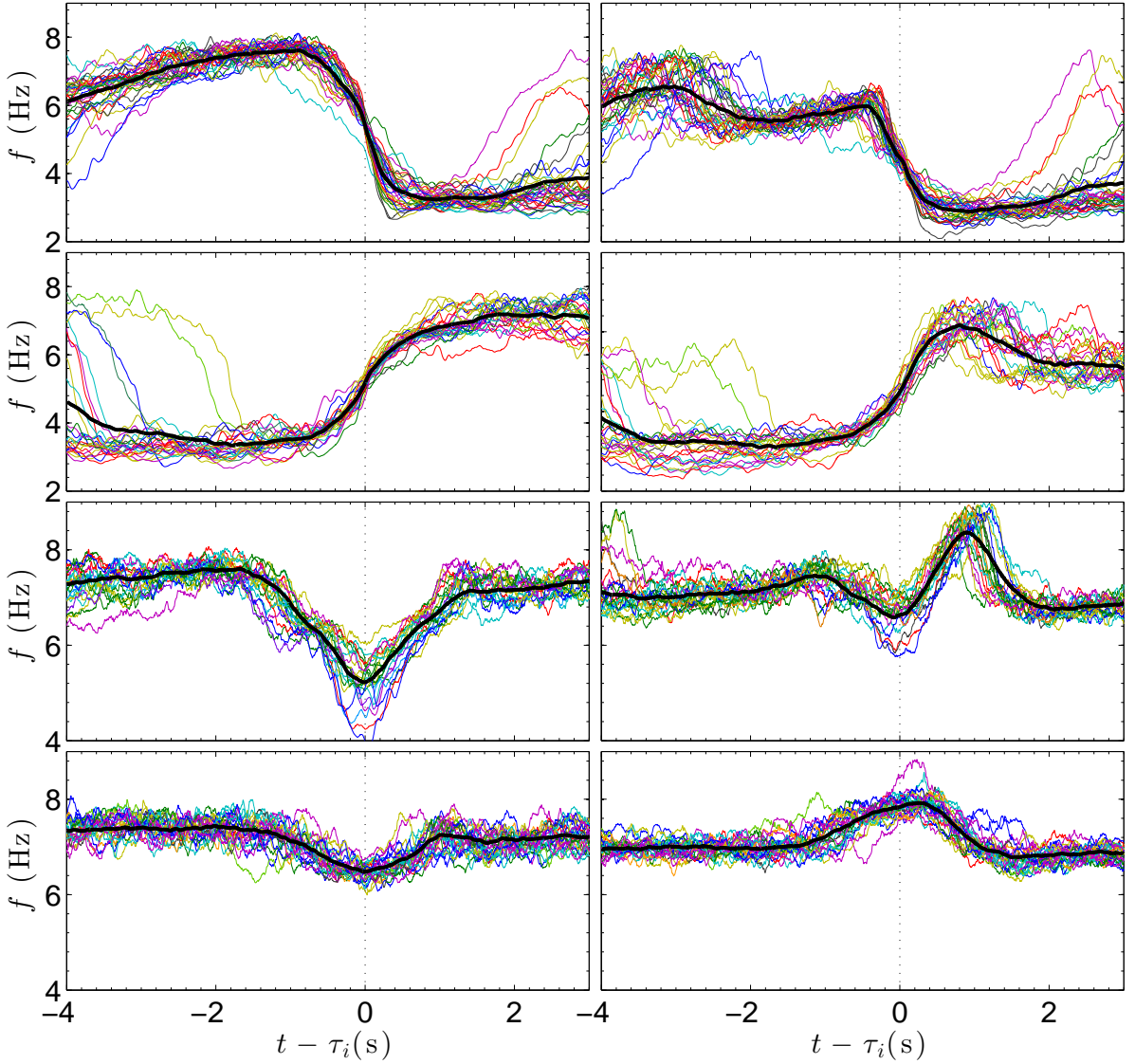}
	\caption{The four types of events observed in our experiment. Left panels, most-forced impeller temporal recording (therefore, top impeller for $\gamma > 0$ and bottom impeller for $\gamma < 0$), and right, least-forced impeller. From top to bottom: transitions from $(s)$ to $(b_2)$ (thin lines) for $\gamma = -0.089$, and mean profile of the 195 events (thick black line); transitions from $(b_2)$ to $(s)$ (thin lines), and mean profile (thick black line) for the same experiment; big excursions between $(s)$, $(i_2)$ and $(b_2)$ for a $\gamma = -0.067$, and mean profile for the 63 events; small excursions between $(s)$ and $(i_1)$ for an experiment at $\gamma = 0.019$, with 79 reported events. Only 25 of the events, selected at random, have been plotted.}
	\label{fig:allevents}
\end{figure}

Even though the flow is in a turbulent regimes, with many degrees of freedom, the transitions in between states observed in \fref{fig:tempseries} are reminiscent of low-dimensional dynamics. To verify this assertion, we may extract mean transition and excursion profiles from our measurements. To do so, the temporal recordings of the impeller speeds have been filtered to $1$~Hz. An algorithm is then used to detect either local extrema or inflection points of the most-forced impeller to respectively detect excursions, as seen in \fref{fig:tempseries}$b$ and \ref{fig:tempseries}$c$), and transitions, as seen in \fref{fig:tempseries}$d$ and \ref{fig:tempseries}$e$). The times $\tau_i$ of these events can be used to superpose all the events from both impellers (see \fref{fig:allevents} in order to verify the quality of such a superposition), and to build an average transition profile (represented by the thick black lines of \fref{fig:allevents}). From this study, two conclusions can be deduced: the general transition and excursion scenario is always the same, the superposition of all events being particularly effective for the transitions, with a very representative average transition profile defining a preferred ``path" characteristic of low dimensional systems subject to noise. This assumption holds for the excursions, where a larger dispersion might be observed on the superposed signals. Finally, we notice that the $(s) \to (b_2)$ mean transition profile is very different from the $(b_2) \to (s)$ profile: the transitions path from $(s) \to (b_2)$ and $(b_2) \to (s)$ are not symmetric to each other.

This situation is therefore very different from magnetic field reversals that have been observed in similar conditions (comparable Reynolds and $\theta$), using a swirling flow of liquid metal under asymmetric, $\theta \neq 0$ forcing conditions~\cite{Berhanu2007}. In such magnetic case, the transition path from positive to negative magnetic field value are symmetric of the transition path from negative to positive, a behaviour that can be reproduced by a model of two coupled stochastic equations~\cite{Petrelis2008}. Therefore, even if our hydrodynamical experiment shows signatures of low dimensional systems, it may be necessary to include more than two coupled stochastic equation to model its behaviour. 

\subsubsection{Distribution of residence time}

\begin{figure}
	\centering
	\includegraphics[width = \textwidth]{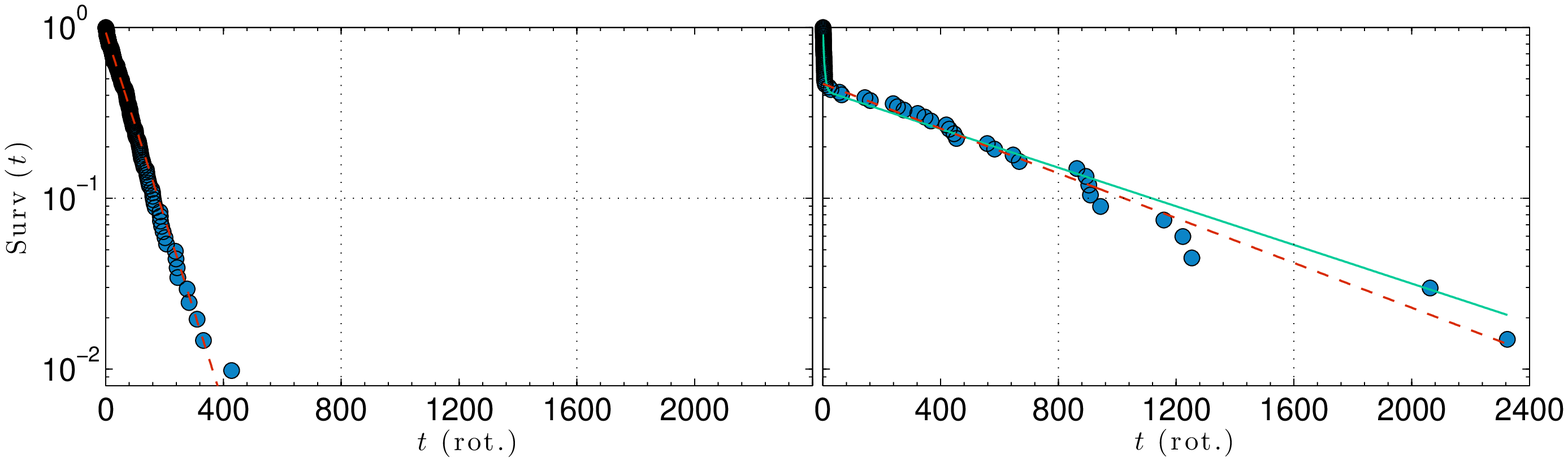}
	\caption{Survival distribution (see \eref{eq:survival}) of the residence time (expressed as multiples of mean impeller rotation time) in the slow state $(b_1)$ for unsteady experiments with approximately $100$ ``rare" events. Blue circles, experimental data; red dashed line, simple exponential fit; solid line, exponential fit with two characteristic times. Left, experiment for $\gamma = -0.094$. Right, experiment for $\gamma = 0.049$.}
	\label{fig:surv}
\end{figure}

Experiments with rare ``bursts" of impeller speeds (\fref{fig:tempseries}$e$) reveal large variations of the residence times in the slow state between bursts. For such experiments, increasing $\gamma$ is equivalent to increase the characteristic time between events, for it naturally diverges at $\gamma$ values corresponding to the steady $(b)$ branch. However, the distribution of times around this characteristic value can indicate the origin of such bursts: reversals of the flow with an exponential distribution have been outlined for a similar von K\'arm\'an experiment~\cite{Torre2007}, allowing an interpretation of this multi-stability once again in terms of stochastic processes, more specifically Kramers escape time processes~\cite{Kramers1940}. The quasi-steady states observed in~\cite{Torre2007} can thus be seen as finite-size potential wells between which a particle with noise may alternatively jump. A particularly convenient way of characterising exponential distributions relies on survival functions, defined as inverse cumulative functions:
\begin{equation}
\label{eq:survival}
{\rm Surv} (t) = 1 - \int_0^t p(t') {\rm d} t'
\end{equation}
where $p$ is the p.d.f. of the residence time. The semi-logarithmic plot of survival functions therefore shows a straight line for any exponential distribution and does not require the use of bins. \Fref{fig:surv} displays the two typical residence time distributions observed in our experiments: while the first one fits the model almost perfectly, the other one ---~generally observed at asymmetries very close to the steady branch $(b)$~--- exhibits another distribution with two characteristic times, a ``short" time and a ``long" time. Such distributions are surprising, knowing that all transitions have the same characteristic shape, and should therefore not make any distinction between ``short-life" and ``long-life" transitions.

\section{Discussion}

This article describes the influence of the Reynolds number and the control nature on a hysteresis cycle originating from a global bifurcation of statistically steady states in a turbulent flow, as previously studied in~\cite{Ravelet2004}. Both of these ``control parameters" affect the hysteresis cycle, but in two different manners.

\subsection{Influence of the Reynolds number on the hysteresis cycle}

Section~\ref{hyst_evol_re} examines the influence, in speed control (described in~\cite{Ravelet2004}), of the \emph{shape} of the hysteresis cycle with the Reynolds number. Whereas the velocity measurements performed in~\cite{Ravelet2008} indicated that the transition to turbulence was attained for ${\rm Re = 3300}$, the apparition of the memory effect (or hysteresis) in the experiment arises for higher Reynolds numbers, $5000 \leq {\rm Re} \leq 6700$. The apparition of such an effect results from two processes: the apparition of local discontinuities of the reduced torque difference $\Delta \gamma$, creating five continuous branches of steady states which progressively grow until they reach $\theta = 0$, a necessary condition to be visible on the figure of~\cite{Ravelet2008}.

The main characteristics of the cycle (width and height), evaluated at various Reynolds numbers, reveal power laws dependence in $| 1 / \log ({\rm Re}) - 1/\log({\rm Re_c}) |$ both for $\Delta \gamma$ and $\Delta \theta$ reminiscent of the susceptibility divergence observed in~\cite{Cortet2011} in the other direction of rotation, $(+)$ also reminiscent of critical phenomena. However, the thresholds in $\Delta \gamma$ and $\Delta \theta$ being different, it is difficult to link the apparition of the hysteresis cycle with only one bifurcation (or transition).

\subsection{Influence of the control nature on the steady states}

Section~\ref{two_controls} reports the effect of the nature of the forcing on the hysteresis cycle for high Reynolds numbers. The forcing nature has only limited effect for control parameters accessible to both speed and torque-imposed experiments: both controls exhibit the same steady states with very similar average velocity fields. The main difference between the two controls rather resides in the set of accessible experimental conditions: torque control experiments may be performed in a speed control ``forbidden zone". Such torque control experiments are no longer steady, revealing multiple peaks of the conjugate asymmetry response ---~$\theta$~--- associated with three quasi-steady states described using stereoscopic PIV. Two of them have been defined by continuity of the speed-control steady states, $(b)$ and $(s)$, but a third quasi-steady state ---~$(i)$~--- emerges, is able to last up to 100 impeller rotations, and does not correspond to a linear combination of $(b)$ and $(s)$ steady states. Preliminary experiments using Laser Doppler Anemometry are currently being conducted to question the origin of such unsteady regimes. Inertial-scale sensitivity to the forcing type could indeed break Kolmogorov's hypothesis of universality of the inertial range.

Additionally, the dynamics between these steady states is akin to low-dimensional models of stochastic processes, with excursions and transition between quasi-steady states following reproducible, well-defined paths which can be viewed as extremal trajectories in an unknown potential energy surface. Furthermore, experiments conducted in the ``rare event" region reveals exponential distribution of residence time in the slow $(b)$ state: transitions from $(b)$ to $(i)$ and $(s)$ therefore result from a random process with no memory, as would a simple Kramers escape time process~\cite{Kramers1940}.

\subsubsection{A very simple model of the flow}

It is tempting to draw an analogy between the von K\'arm\'an flow and an electrical system. Our experiment is subject to a permanent flux of kinetic angular momentum, which is injected at a rate $C_1$ ---because of the kinetic momentum theorem--- by the bottom impeller and transported by the flow to the top impeller where it is absorbed at a rate $C_2$~\cite{Marie_Daviaud}. The difference $C_1-C_2$ ``leaks'' through the side wall boundary layer. Kinetic angular momentum cannot be dissipated and cannot diverge in the bulk of the flow: hence, this quantity behaves like the total number of charges $q$ in an electrical circuit.

In contrast, the kinetic energy, the usual hydrodynamical quantity, is not conserved in our experiment. The von K\'arm\'an flow dissipates mechanical energy very efficiently, for it is introduced at a rate $P= 2 \pi (C_1 f_1 + C_2 f_2)$ by the motors. In this picture, the $z$-component of the rotation vector of the impellers, $2 \pi f_1$ and $-2 \pi f_2$, serves as analogue of the electrical potential. The cylinder wall, which has a $0$ rotation rate, plays the role of the electrical ground.

\begin{figure}
\begin{center}
	\includegraphics[clip,width=0.65 \textwidth]{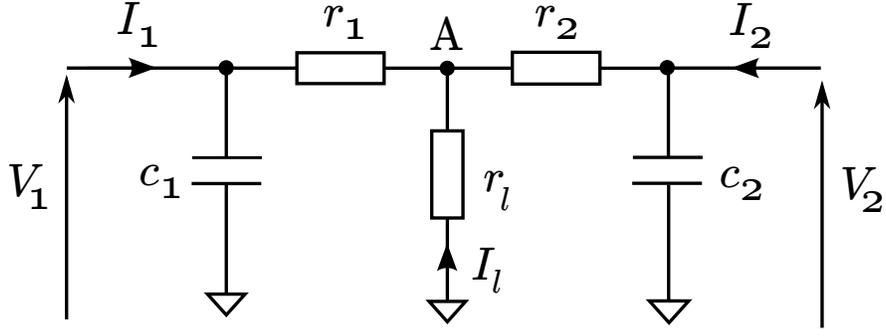}
	\caption{\label{Figures:schema} Sketch of an electronic circuit, analogue of the von K\'arm\'an flow.}
\end{center}
\end{figure}

In torque control, our experiment can therefore be modelled as an electrical circuit which can develop electrical potential instability for forced currents $I_1$ and $I_2$ (the equivalents of the applied torques $C_1$ and $C_2$) through two of its poles (the third one being tied to ground). We have presented in \fref{Figures:schema} a sketch of this circuit. The left pole (the analogue of the bottom impeller) permanently injects a current $I_1$, which is extracted at a rate $I_2$ at the right pole (or, equivalently, the top impeller). Two capacitors of capacities $c_1$ and $c_2$ represent the ability of the fluid next to the impellers to store angular momentum and allow the rotation rates of the impellers to increase. The two resistors of resistances $r_1$ and $r_2$ represent the capacity of the flow to transport angular momentum away from turbine 1 and towards turbine 2. Finally, a last resistor $r_l$ represents the leak of angular momentum to the cylinder wall. 

In the permanent regime, $I_{l} = I_1-I_2$ plays the role of the torque difference provided by the impellers on the flow, $C \gamma$. Close to the permanent regime with $\gamma=0$, $V_{A} \approx  (V_1 + V_2)/2$ can be seen as the difference of impeller speeds $f \theta$. The actual ``characteristic curve" of all the dipoles in our model is obviously not known and is at the core of the understanding of the problem of fluid mechanics. In the following, we will only perform a small-signal analysis of the circuit in which we will only consider small perturbations of the flow around a working point (we assume that at least one working point exists), for which we will consider the responses to be linear. We will denote the small-signal quantities as the lower-case equivalent of their permanent regime counterparts. We will consider perturbations verifying $i_1 = 0$ and $i_2 = 0$: the external sources are not affected by our perturbations (in our hydrodynamical experiment, this corresponds to a perfect torque regulation).
Decomposing the perturbations in the Fourier space at pulsation $\omega$, one obtains:
\begin{eqnarray}
v_A &=& (1 + j r_1 c_1 \omega) v_1 \\
v_A &=& (1 + j r_2 c_2 \omega) v_2 \\
v_A &=& \frac{v_1 / r_1 + v_2 / r_2}{1 / r_1 + 1/ r_2 + 1/r_l}
\end{eqnarray}
The system is overdetermined: it can nevertheless be solved $\omega$ satisfies the dispersion relation:
\begin{eqnarray}
\label{eq:expr_omega}
 \frac{1}{r_1} + \frac{1}{r_2} + \frac{1}{r_l}&=& \frac{1}{r_1 (1 +  j r_1 c_1 \omega)} + \frac{1}{r_2 (1 + j r_2 c_2 \omega)}
\end{eqnarray}
We will consider the case of low-frequency perturbations, which verify $| r_1 c_1 \omega | << 1$ and $| r_2 c_2 \omega | << 1$. The first-order solution of equation~\ref{eq:expr_omega} then reads:
\begin{equation}
\omega = \frac{j}{r_l (c_1 + c_2)}
\end{equation}
Infinitesimal perturbations of $v_A$ will therefore behave as:
\begin{equation}
v_A(t) \propto \exp \left ( - \frac{t}{r_l (c_1 + c_2)}  \right ) 
\end{equation}
Such perturbations are stable unless the leak resistance $r_l$ is negative. This condition can be verified in a \emph{differential} sense if $\partial V_A / \partial I_l$ is negative around the chosen working state. Following our analogy, an equivalent condition in our turbulent flow is :
\begin{equation}
\aleph = \partial \theta / \partial \gamma < 0
\end{equation}
which is the case in average and in the intermediate branch $(i)$. Hence, perturbations grow until non-linear terms of the equation become significant, leading to ``spontaneous" jumps in the hydrodynamical experiment. The symmetry of this perturbation implies that $v_1$ and $v_2$ are of the same sign: thus, the equivalent mechanical perturbation will be antisymmetric with a possible delay between impellers depending on the relative values of $r_1 c_1$ and $r_2 c_2$. This model is certainly too simplified, since it cannot take into account the inherent turbulent out-of-equilibrium noise, but it still captures several key features of the actual experiment.

\subsubsection{Link with other negative differential responses}

In the previous section, our results assumed that negative differential responses existed in electrical components around a working point $(V_A, I_l)$. Such components exist: tunnel diodes~\cite{Esaki1958,Ridley1963} exhibit negative differential resistances. This phenomenon is also observed in complex fluids, for example in Taylor-Couette geometries~\cite{Bonn1998,Wilkins2006} where the differential response to shear can become negative. All these systems can work under conjugate conditions: 
\begin{itemize}
	\item von K\'arm\'an flows can impose impeller speed asymmetry $\theta$ or shaft torque $\gamma$.
	\item electrical dipoles can be examined under imposed voltage $u$ or current $i$.
	\item Taylor-Couette flows can either impose cylinder speed or torque. Imposing the cylinder speed difference $\Delta v = v_{\rm in} - v_{\rm out}$ is analogous to impose the shear rate $\dot{\epsilon}$ and fixing the impeller torque constrains the shear stress $\tau$.
\end{itemize}
Therefore, von K\'arm\'an torque control, electrical dipoles under imposed current and Taylor-Couette experiments with constrained shear rate should yield similar results in our description. The experiment of Bonn \emph{et al.}~\cite{Bonn1998} and the current controlled negative differential resistances, studied by~\cite{Ridley1963}, are closest to our experiment: in addition of displaying negative differential responses of the same nature as our flow, they are ``bulk" experiments in which spatial instabilities can be observed. In contrast with electrical components and our von K\'arm\'an flow, none of the two control parameters in the Taylor-Couette flow is a flux of a conserved quantity which cannot diverge in the bulk. Hence, the electrical analogy is the most relevant.

In all cases, the product of the two controls is closely linked to the definition of the injected power $\mathcal{P}$ to the experiment. For a von K\'arm\'an experiment, we have:
\begin{eqnarray}
\label{eq:power}
\mathcal{P} &=& 2 \pi (f_1 C_1 + f_2 C_2)    \\
			&=&	2 \pi f C (1 + \gamma \theta) 
\end{eqnarray}
The injected power involves, in this particular case, $\theta \gamma$, which varies respectively as a function of $\theta$ in speed control, or $\gamma$ in torque control. Still, the shape of the hysteresis cycle is only weakly dependent on the values of the Reynolds number in this range of experiments, and nothing (at least theoretically) prevents from performing experiments at fixed $f C$ to examine the dependence of $\mathcal{P}$ as a function of $\theta$, in a modified speed control, and $\gamma$, in a modified torque control. In this case, the injected power $\mathcal{P}$ is, up to some additive constant, defined by the product $\gamma \theta$. This relation is much simpler in the case of the electrical dipole, for which the simple relation:
\begin{equation}
\mathcal{P} = u i
\end{equation}
links the two imposed quantities. Eventually, for Taylor-Couette geometries, the initial relation is somewhat the same as for a von K\'arm\'an flow (defined at \eref{eq:power}) but with a different final result. It can be shown, in the case where only one cylinder is rotating, and assuming that the shear stress does not depend on $r$, that :
\begin{eqnarray}
\mathcal{P} &=       & 2 \pi (f_1 C_1 ) \\
			&\propto & \tau \int_{r_{\rm in}}^{r_{\rm ext}} \dot{\epsilon} ( r )~{\rm d} r
\end{eqnarray}
which is again the product of the ``conjugate" imposed quantities.

The negative differential responses of~\cite{Bonn1998} are observed imposing the shear rate, and are associated with a hysteresis cycle when the shear stress is imposed with sudden jumps of the shear rate. In addition, the steady states near the jumps tend to display finite lifetime, as do our experiments near the edge of the $(s)$ and $(b)$ branches in speed control~\cite{Ravelet2005}. The negative differential zone is in this case associated with a shear banding transition. The electrical dipoles of~\cite{Ridley1963} have been predicted to display hysteresis for imposed voltage, vanishing at imposed current: the negative differential resistances are in that case associated with unsteady filaments of high current surrounded by low-current regions. Such filaments in a von K\'arm\'an experiment could result in the presence of portions of the fluid rotating with a different angular speed than the rest of the fluid. Bifurcated $(b)$ quasi-steady states might show such domains, two tori of opposite global rotation being observed in these cases. Hence, our flow and these bulk dipoles have many common points, the main difference between them being the noise level and nature.

\subsubsection{Possible link with ensemble inequivalence}

Classical thermodynamics states, with ensemble equivalence, that a system with fixed energy density ---~thus, described in the \emph{microcanonical} ensemble~--- will behave the same manner as a system at imposed temperature ---~ in the \emph{canonical} ensemble: in particular, the set of thermodynamically stable states are the same in both ensembles. This result assumes that the (two-particle) potential interaction possesses a \emph{short-range} property, decreasing at least with a power law of exponent greater than the system dimensionality. Long-range interacting systems, in contrast, do not respect this condition and lead to violation of statistical ensemble equivalence~\cite{Dauxois2000,Barre2001}. This section will first explain the origin of such a difference, to evaluate afterwards the relevance of the analogy for a von K\'arm\'an turbulent flow.


In thermodynamics, the condition of positivity of the specific heat is linked to a general property of the entropy:
\begin{eqnarray}
	\label{eq:cv}
			 c_v &=& \left ( \frac{\partial e}{\partial T}\right ) _V  \\
	\frac{1}{c_v}&=& - T^2 \left( \frac{\partial^2 s}{\partial e^2} \right)_V
\end{eqnarray}
The entropy $s$ is generally considered an increasing, concave function of the internal energy $e$, imposing a positive specific heat. In canonical ensemble $c_v$ is also linked to the fluctuations of the average energy density, namely:
\begin{equation}
c_v = \langle e^2 \rangle - \langle e \rangle^2 \geq 0
\end{equation}
Still, convex intruders in the entropy-energy relation might exist. Assuming additivity of the entropy $S$ and internal energy $U$, such intruders spontaneously generate first-order phase transitions, implying that specific heats remain positive in both ensembles. Long-range interacting systems, in contrast, can sustain negative specific heats, providing distinct sets of stable equilibrium positions in the two ensembles. Interestingly, the very definition of $c_v$, linking in a partial derivative $e$, fixed in microcanonical ensemble, and $T$, fixed in canonical ensemble, is once again a (possibly negative) differential response. 

Changing the nature of the command in our von K\'arm\'an experiment could represent, in this picture, an ensemble switching between micro-canonical and canonical ensembles, the former being played by torque control, and the latter by speed control, assuming that the underlying physics of our von K\'arm\'an flow involves some sort of long-range interactions. Evidence supporting this assumption can be found in two-dimensional~\cite{Chavanis2002} an more recently in three dimensional axisymmetric flows~\cite{Thalabard2013}. A major difference limits, though, such a comparison: there is currently no variational formulation of turbulence from which extremal quantities could be derived. This absence is probably originating from the out-of-equilibrium nature of turbulence, which is another aspect not accounted for in this analogy: the multi-stability observed inside our hysteresis cycle cannot be accounted for in an equilibrium description. 

\subsubsection{Comparison with turbulent Rayleigh-B\'enard convection}

\ADD{The duality between controls, observed here in a turbulent flow, might also be present in another turbulent classical experiment, namely Rayleigh-B\'enard convection. In the turbulent regime, Rayleigh-B\'enard convection exhibits hysteresis of the response --~the Nusselt number, quantifying the importance of convective heat flux through the cell~-- for very large values of the control parameter --~the Rayleigh number, expressing the non-dimensional temperature difference between the boundary plates~--~\cite{Ahlers2009}. Reversals and cessation of the turbulent large-scale wind have already been observed both in two and three-dimensional turbulent convection experiments and numerical simulations (see, for example,~\cite{Brown2006, Sugiyama2010} for details). They seem to be good candidates for counterparts of the hysteretic states, in the same manner as our up and down transitions are counterparts of the steady $(s)$ and $(b)$ hysteretic states in the von K\'arm\'an flow. However, such events have, according to the existing literature, almost no impact on the Nusselt number for a fixed Rayleigh number; the literature rather attribute the difference between the branches to changes in the nature of the boundary layer~\cite{Ahlers2009a}. In addition, all these Rayleigh-B\'enard experiments have been conducted with the same boundary conditions: smooth, perfectly conducting plates (imposed Rayleigh number). Switching from ``speed" to ``torque" control in Rayleigh-B\'enard experiments would require to impose the conjugate quantity of the temperature gradient $\Delta T$ of the cell with respect to the injected power $\mathcal{P}$, in other words an imposed entropy flux injected at the boundary plates. Rayleigh-B\'enard convection has been studied for imposed heat-flux --~or Nusselt number~--, an approximation of imposed entropy flux, but experimental results are scarce in this domain: approaching experiments with poorly conducting plates~\cite{Roche2005} show neither boundary-layer multistability nor difference with well-conducting copper plates. Our results call for more experiments in this domain, especially since rough boundary conditions have been reported to promote multi-stability of the boundary layer affecting the Nusselt number in recent experiments~\cite{Salort2014}.}

\section{Conclusion}

In this article, we have studied the continuous evolution of the response $\gamma$ to an imposed symmetry-breaking field $\theta$ with the Reynolds number, below and above the transition to turbulence. The particular shape of the impellers progressively creates torque discontinuities growing to become progressively hysteretic. Interestingly, up to five branches might coexist, some of which disappear for Reynolds numbers above the critical Reynolds number of the onset of turbulence (identified as the onset of a Kolmogorov spectrum). The main characteristics of the cycles (width and height) respectively decrease and increase monotonically with ${\rm Re}$ with possible power-law behaviour.

In addition, we have discussed the influence of the nature of the forcing on the hysteresis cycle for high Reynolds numbers: in contrast with speed control, \emph{torque} control experiments connect the three hysteretic branches continuously, displaying negative differential responses $\partial \theta / \partial \gamma$, closely linked to a multi-stability emerging in the experiment, as evidenced by an electrical toy-model of the flow. This dynamics can be seen as a time counterpart of the spatial heterogeneities observed in other out-of-equilibrium systems where the noise has thermal origins (shear banding fluids and bulk solid state electrical components), which also display negative differential responses. 

Ensemble inequivalence results may explain why the negative responses observed in our flow can be sustained, assuming that the change of the control is equivalent to switching between statistical ensembles. In this framework, microcanonical systems may exhibit intermittent dynamics~\cite{Torcini1999} or spatial heterogeneities~\cite{Mori2010}. 

\ADD{Two approaches are possible to compare our results to other turbulent flows. First, it is possible to find other turbulent flows where conjugate controls can be managed. A direct comparison of our results with turbulent Rayleigh B\'enard convection ---~where hysteresis between multiple turbulent states has been observed at large Rayleigh numbers~--- would require experiments at imposed Nusselt number. To the best of our knowledge, such experiments are uncommon~\cite{Roche2005} and have not evidenced any type of multi-stability. Such multi-stability has however been observed at very high Rayleigh numbers for rough boundary plates~\cite{Salort2014} ; it is obviously very difficult to express the difference between rough and smooth plates in terms of our ``conjugate" forcing framework. Pipe flows also allows conjugate controls : experiments at imposed flow-rate have been conducted, and have neither reported~\cite{Darbyshire1995} any difference with the classical case of pressure control.}

\ADD{Another way to compare our results with other, more loosely connected turbulent flows, would be to investigate the small scale evidence of the large-scale (and forcing) fluctuations in the experiment. These fluctuations are already known to challenge the ``universal" nature of the Kolmogorov constants in the inertial range~\cite{Chien2013}. More specifically, small-scale signatures of our multi-stability may be found both in our experiment and other experiments displaying less obvious forms of hysteresis and multi-stability for varying boundary conditions. This could be the case in Rayleigh-B\'enard convection (see~\cite{Salort2014}) or also for flows past hydrofoils where the hysteresis cycle of stall vanishes, being replaced by a multi-stable region, for thicker hydrofoils~\cite{Sarraf2007}.}

\section*{Acknowledgements}

We are greatly indebted to C. Wiertel Gasquet and V. Padilla for technical and mechanical assistance during the experiments. We are thankful for G. Mancel for sharing experimental data.

\section*{References}

\bibliographystyle{unsrt}
\bibliography{biblio}

\end{document}